\documentclass[a4paper,10pt,twoside]{cpc-hepnp}
\usepackage{CJK,upgreek,fancyhdr}
\usepackage{multicol}
\usepackage{graphicx}
\usepackage{booktabs}
\usepackage{amssymb,bm,mathrsfs,bbm,amscd}
\usepackage[tbtags]{amsmath}
\usepackage{lastpage}

\newcommand{\bfg}{\begin{figure}}
\newcommand{\efg}{\end{figure}}
\newcommand{\bitm}{\begin{itemize}}
\newcommand{\eitm}{\end{itemize}}
\newcommand{\bnum}{\begin{enumerate}}
\newcommand{\enum}{\end{enumerate}}
\newcommand{\btbl}{\begin{table}}
\newcommand{\etbl}{\end{table}}
\newcommand{\btbu}{\begin{tabular}}
\newcommand{\etbu}{\end{tabular}}
\newcommand{\bcl}{\begin{center}}
\newcommand{\ecl}{\end{center}}

\newcommand{\beq}{\begin{equation}}
\newcommand{\eeq}{\end{equation}}
\newcommand{\beqr}{\begin{eqnarray}}
\newcommand{\eeqr}{\end{eqnarray}}
\begin{document}




\title{Jet shape modifications at the LHC energies by JEWEL
}

\author{%
      WAN Ren-Zhuo$^{1;1)}$\email{wanrz@wtu.edu.cn}%
      \quad DING Lei$^{1}$
      \quad GUI Xi$^{1}$
\quad YANG Fan$^{1}$
\quad Li Shuang$^{2;2)}$\email{lish@ctgu.edu.cn}%
\quad ZHOU Dai-Cui$^{3;3)}$\email{dczhou@mail.ccnu.edu.cn}%
}
\maketitle
\vspace{0.2cm}

\address{%
$^1$ Nano optical material and storage device research center, School of electronic and electrical engineering, Wuhan Textile University, Wuhan 430200, China\\
$^2$ College of Science, China Three Gorges University, Yichang 443002, China \\
$^3$ Key Laboratory of Quark and Lepton Physics (MOE) and Institute of Particle Physics, Central China Normal University, Wuhan 430079, China\\
}

\begin{abstract}
Jet shape measurements are strongly suggested to explore the microscopic evolution mechanism of parton-medium interaction in ultra-relativistic heavy-ion collisions. In this paper, jet shape modifications are quantified by fragmentation function $F(z)$, relative momentum $p_{T}^{rel}$, density of charged particles $\rho(r)$, jet angularity $girth$, jet momentum dispersion $p_{T}^{Disp}$ and $LeSub$ for proton-proton collisions at 900 GeV, 2.76 TeV, 5.02 TeV, 7 TeV and 13 TeV, as well as for lead-lead collisions at 2.76 TeV and 5.02 TeV by JEWEL. A differential jet shape parameter $D_{girth}$ is proposed and studied at smaller-radius jet $r<0.3$. The results indicate that medium effect is dominant for jet shape modifications, which has weak dependence on center of mass energy. The jet fragmentation is enhanced significantly at very low $z<0.02$ and fragmented jet constituents are spread to larger jet-radius linearly for $p_{T}^{rel}<1$. The waveform attenuate phenomena is observed in $p_{T}^{rel}$, $girth$ and $D_{girth}$ distributions. The comparison results on $D_{girth}$ from $pp$ to $Pb+Pb$ where the wave-like distribution in $pp$ collision is ahead of $Pb+Pb$ collisions in small jet-radius intervals, is interesting to hint the medium effect.

\end{abstract}



\begin{multicols}{2}
\section{Introduction}
    One goal of jet physics is to explore the microscopic properties of hot-dense QCD matter, quark-gluon plasma (QGP), created in ultra-relativistic heavy-ion collisions \cite{QGP,jetphy,jetquenching}. Jets produced by partons in early stage from collide nucleus will travel through QCD matter and carry multi-scale physics evolution information. The interplay between elementary scattering, subsequent branching process and strongly coupled parton-medium interactions can lead to modification of jet shapes with respect to jet fragmentation in the absence of medium. The jet structure modifications from proton-proton collisions to heavy-ion collisions could be investigated by a set of jet shape quantities and will shed light on understanding of jet energy loss mechanism in medium, color coherence and fundamental medium properties.

    The latest results at the LHC on charged particle nuclear modification factor measurement at a wide momentum range show a sectionalized behavior giving rise to a clear soft and hard pQCD region \cite{raa}. The observations of more highly unbalance di-jet with increasing event centrality \cite{dijet1}, the suppression of inclusive jet yield by about a factor of two in central heavy-ion collisions relative to peripheral collisions, as well as the correlation of the jet suppression with missing transverse momentum \cite{dijet2,alicejetsuppression,plb2013}, indicate that jets energy redistribute within jet in medium relative to vacuum and is non-ignorable fraction of jet energy loss especially at large angles relative to the jet axis. The fragmentation yield is found a reduction at intermediate $z$ and enhancement at small $z$ in central collisions relative to peripheral collisions \cite{plb2014, epjc2017}. The azimuthal di-hadron correlation excess yield is found more pronounced in sub-leading jet predominantly from several to 20 GeV arising from soft particles \cite{corellation}. In \cite{smallrALICE}, the jet shapes modifications are studied at small jet cone size ($r<0.2$) and found an inconsistent with a fully coherent energy loss picture. Those results supply rich inputs for theoretical calculations and phenomena modeling on jet-medium interactions.

    From theoretical side, most of phenomena studies on jet energy loss are based on pQCD, which could explain data well at intermediate momentum range, but incapability to cover an entire kinematic range in non-pQCD region \cite{quenchingweight,qiopp}. One crucial reason is the treatment of gluon radiation by supposing collinear radiation and model dependent on momentum exchange among the medium and parton dynamical evolution for leading order or next leading order contribution, especially for central $Pb+Pb$ collisions\cite{qgpbrick}. Besides, several physics mechanisms are coexistence and competition when partons traverse within medium, for instance recoil and non-recoil, coherence and de-coherence effects, depending on QCD local equilibrium temperature, medium density, path length, gluon formation time and momentum exchange etc., which redistribute the parton energy at a relative larger angle with respect to its initial directions \cite{wan}.

    Jet-medium interactions cover full kinematics range and include perturbative and non-perturbative effects at the LHC energies. It is essential to study the jet shape modification at various collision systems by a large set of jet shape quantities. In this paper, the measurement of jet shape observable, for instance fragmentation function $F(z)$, relative momentum $p_{T}^{rel}$, density of charged particles $\rho(r)$, jet angularity $girth$, jet momentum dispersion $p_{T}^{Disp}$ and the difference between leading and sub-leading constituent transverse momentum $LeSub$ for proton-proton collisions at 900 GeV, 2.76 TeV, 5.02 TeV, 7 TeV and 13 TeV, as well as for lead-lead collisions at 2.76 TeV and 5.02 TeV, are systematically studied by JEWEL \cite{jewel}. One additional differential jet shape parameter of $D_{girth}$ is proposed and studied at $r<0.3$ to investigate the jet shape evolution.

\section{Jet shape measurements}

    Jet shape measurements have been suggested in exploring the microscopic evolution mechanism of parton-medium interaction in ultra-relativistic heavy-ion collisions. For different physics motivations, several jet shape observations have been proposed using jet constituents, overall jet-by-jet quantities and jet cluster historical information. For instance, fragmentation function, missing $p_{T}$ method and jet-track angular correlations measurements by using intra or inter-jet distributions are dedicated to investigating the longitudinal share of energy within jet and large angle radiations. The overall jet shape observations built on jet-by-jet function of jet constituent 4-momenta, such as jet mass, jet granularity and jet momentum dispersion etc., could probe the jet energy loss dependence on large angle soft particle emission. Jet grooming measurements based on jet cluster historical information could locate the splitting phase space where  the medium-induced effects are expected \cite{groom1,groom2}. This paper presents a jet shape study in $Pb+Pb$ and $pp$ collisions by inclusive constituents of jet and jet-by-jet quantity to anchor jet evolution and energy loss jet within QCD matter.

    In \cite{epjc2011}, ATLAS presents the fragmentation function and transverse profiles in $\sqrt{s}$=7~TeV $pp$ collisions at a wide jet momentum range of 25$~$GeV/c$<p_{T,jet}<$500$~$GeV/c and found discrepancies between various models and data. Fragmentation functions show a reduction yield at $0.04\leq z \leq0.2$ and an enhancement for $z\leq0.04$ in $\sqrt{s_{NN}}$=2.76 TeV $Pb+Pb$ collisions by ATLAS \cite{plb2014}. It was also measured in ALICE \cite{arXiv181109742} from di-hadron correlations in $\sqrt{s}$=7 TeV $pp$ and $\sqrt{s_{NN}}$=5.02~TeV $p+Pb$ collisions and didn't observe significant cold nuclear matter effect. In \cite{smallrALICE,jhep2016}, the jet shape modifications were studied in small-radius jets to discriminate the relative quark and gluon jet fractions, which also suggested that the medium was able to resolve the jet structure at angular scales smaller than $r = 0.2$. However, detail and systematic study are highly required to quantify the gluon induced fragmentation modifications, better constrain energy loss models and isolate quark and gluon jet fraction at intermediated momentum range for multiple collision systems. In this paper, the below observations are used.

    Jet fragmentation function $D_{i}^{h}(z,Q)$ is defined as the probability that a hadron carries longitudinal momentum fraction $z$ of $p_{jet}$ in experiment. $z$ is defined:
 \begin{equation}
 \label{zdef}
 z=\frac{\vec{p}_{jet} \cdot \vec{p}_{ch}}{|\vec{p}_{jet}|^{2}}
 \end{equation}

 \noindent  which connects color partons and colorless hadrons to constrain QCD-motivated models from experiments. The quantity of $F(z, p_{T,jet})$ is measured as:

\begin{equation}
\label{fzdef}
F(z, p_{T,jet})=\frac{1}{N_{jet}} \frac{dN_{ch}}{dz}
\end{equation}

\noindent Where $N_{ch}$ is the number of charged particles in jet and $N_{jet}$ is the number of selected jets to be used for normalization. Two additional quantities $p_{T}^{rel}$ and the density of charged particle $\rho_{ch}$ are also studied. $p_{T}^{rel}$ is the momentum of charged particles in a jet transverse to that jets axis defined as Eq.\,(\ref{ptrel}) and its distribution $f(p_{T}^{rel}, p_{T,jet})$ in Eq.\,(\ref{fptrel}):

\begin{equation}
\label{ptrel}
p_{T}^{rel}= \frac{|\vec{p}_{ch}\times \vec{p}_{jet}|}{|\vec{p}_{jet}|}
\end{equation}

\begin{equation}
\label{fptrel}
f(p_{T}^{rel}, p_{T,jet})=\frac{1}{N_{jet}} \frac{dN_{ch}}{dp_{T}^{rel}}
\end{equation}

\noindent And the density of charged particle $\rho_{ch}$ in $y-\phi$ space is measured as a function of angular distance $r$ of charged particles from jet axis, given by:

\begin{equation}
\label{rhoch}
\rho_{ch}(r) = \frac{1}{N_{jet}} \frac{dN_{ch}}{2\pi rdr}
\end{equation}

  For a complementary study, the jet-by-jet quantities, namely the first radial moment or angularity (or girth), $girth$, the momentum dispersion, $p_{\mathrm{T}}^{Disp}$, and the difference between the leading and sub-leading track transverse momentum, $LeSub$ are also studied. $girth$ and $P_{T}^{Disp}$ are related to the moments of the so-called generalized angularities \cite{girth}. And $LeSub$ is not an IRC-safe observable but indicates robustness against soft background particles contributions. The angularity is defined as
\begin{equation}
\label{eq:angu}
g = \sum_{i \in \rm jet} \frac{p_{\mathrm{T,}i}}{p_{\mathrm{T,jet}}} \Delta R_{\mathrm{jet},i},
\end{equation}

where $p_{\mathrm{T},i}$ is the transverse momentum of the $i$-th constituent and $\Delta R_{\mathrm{jet},i}$ is the distance in ($\eta$, $\varphi$) space between constituent $i$ and the jet axis. This shape is sensitive to the radial energy profile of the jet.

\noindent The momentum dispersion $p_{\mathrm{T}}^{Disp}$ is defined as:

\begin{equation}
\label{eq:ptd}
 p_{\rm T}^{Disp} = \frac{\sqrt{\sum_{i \in \rm jet}  p_{\mathrm{T,}i}^{2}}} {\sum_{i \in \rm jet} {p_{\mathrm{T,}i}}}.
\end{equation}

\noindent $LeSub$ is defined as the difference of the leading  track $p_{\rm T}$ and sub-leading track $p_{\rm T}$,

\begin{equation}
\label{eq:angu}
LeSub = p_{\mathrm{T,track}}^{\rm {lead}}-p_{\mathrm{T,track}}^{\rm {sublead}}.
\end{equation}

For small-$r$ jet, it contributes to isolated pure energy loss effect from other medium effects and correlated background. So the differential jet-by-jet quantities, $D_{girth}$ is proposed, which defined as their distribution of $girth$ inside an annulus of inner radius $r$ and outer radius $r+\delta r$ around the jet axis, which is directly related to the jet energy topological structure and its evolution.

\section{Simulation method}

In this study, JEWEL \cite{jewel} is used to simulate jet production, QCD scale evolution and re-scattering of jets in heavy-ion collisions based on pQCD. It describes the jet evolution and jet-medium interactions simultaneously and dynamically based on leading-order matrix elements plus parton shower method. And all partons belonging to partons showers initiated by hard scattered partons undergo collisions with thermal partons from the medium, leading to both elastic and radiative energy loss. Soft gluon radiation, recoil effect and scattering processes are governed by formation time, which has been shown to agree with analytical calculations in the appropriate limits. LPM effect is included by generalizing the probabilistic formulation in Eikonal limit to general kinematics. For heavy-ion environment, the medium density profile is based on a longitudinally expanding Glauber overlap; the local temperature is sampled to determine the density of scattering centers and their momentum distribution. The initial time $\tau$=0.6 fm and temperature $T_{i}$=0.4 GeV, the critical temperature $T_{C}$=170 MeV are fixed for lead-lead collisions in central collisions. Jets are reconstructed by using all final detectable charged particles and neutral particles with the anti-kt algorithm provided by the FastJet \cite{fastjet} package within $|\eta|<2.0$ and $|\eta|<2.8$ for fragmentation distribution and jet-by-jet quantities measurements in separately. The analysis takes into account the recoil effect for heavy-ion collision and the constituent subtraction method was used to remove the contribution of medium fragments from the parton energy.

\end{multicols}

\begin{center}
\includegraphics[width=1\textwidth]{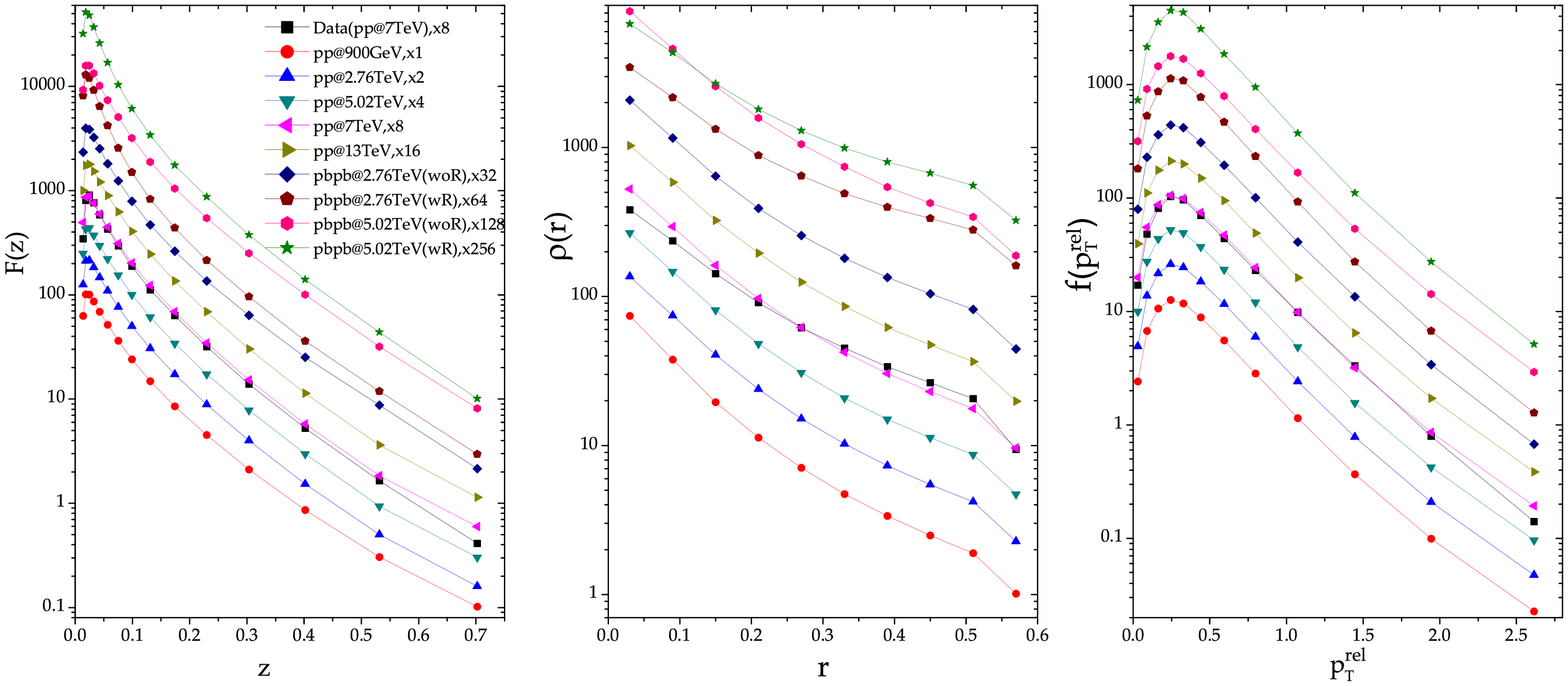}
\figcaption{\label{fig1} $F(z)$, $\rho_{ch}(r)$ and $f(p_{T}^{rel})$ distributions for $pp$ collisions at $\sqrt{s}$=900 GeV, 2.76 TeV, 5.02 TeV, 7 TeV, 13 TeV and $Pb+Pb$ collisions at $\sqrt{s_{NN}}$=2.76 TeV and 5.02 TeV in the range of 25 GeV/c$<p_{T,jet}<$40 GeV/c simulated by JEWEL and compared with data of $pp$ collisions at $\sqrt{s}$=7 TeV\cite{epjc2011}.}
\end{center}
\vspace{-0.3cm}

\begin{multicols}{2}
\section{Results and discussion}

Figure \ref{fig1} present the fragmentation function distributions of $F(z)$, relative momentum $f(p_{T}^{rel})$ and the density of charged particles $\rho_{ch}(r)$ for $pp$ collisions at $\sqrt{s}$=900 GeV, 2.76 TeV, 5.02 TeV, 7 TeV, 13 TeV and $Pb+Pb$ collisions at $\sqrt{s_{NN}}$=2.76 TeV and 5.02 TeV in the range of 25 GeV/c$<p_{T,jet}<$40 GeV/c simulated by JEWEL. The recoil effect is taken into account for $Pb+Pb$ collisions~(referred as $wR$ for recoil effect and $woR$ for without recoil effect). The simulations are compared with data of $pp$ collisions at $\sqrt{s}=$7 TeV. $F(z)$ distributions indicate a strong enhancement at $z<0.1$ and a reduction at $z>0.1$ compared from $pp$ to $Pb+Pb$ collisions with recoil effect. For $pp$ collisions and $Pb+Pb$ collisions without recoil effect, the enhancement or reduction on $F(z)$ depend on collisional energy and participant nucleus.
$\rho(r)$ indicate the same performance that jet constituents spread to large jet-radius with the increase of center of mass energy of collisions, and the medium effect is significant for the modification at relative large angles. $\rho(r)$ distribution is exponential decrease with jet-radius. $f(p_{T}^{rel})$ distribution shows an independent of collisional energy, but strongly modified by QCD medium and present an obvious enhancement at about $0.3<r<1.0$.

To better illustrate the differences between simulations and data, Figure \ref{fig2}, \ref{fig4} and \ref{fig3} show the ratio plots with the reference data of $\sqrt{s}$=7 TeV $pp$ collisions for $F(z)$, $\rho(r)$ and $f(p_{T}^{rel})$ in separately at four jet momentum intervals of [25,40] GeV/c, [40,60] GeV/c, [60,80] GeV/c and [80,110] GeV/c.

\begin{center}
\includegraphics[width=0.5\textwidth]{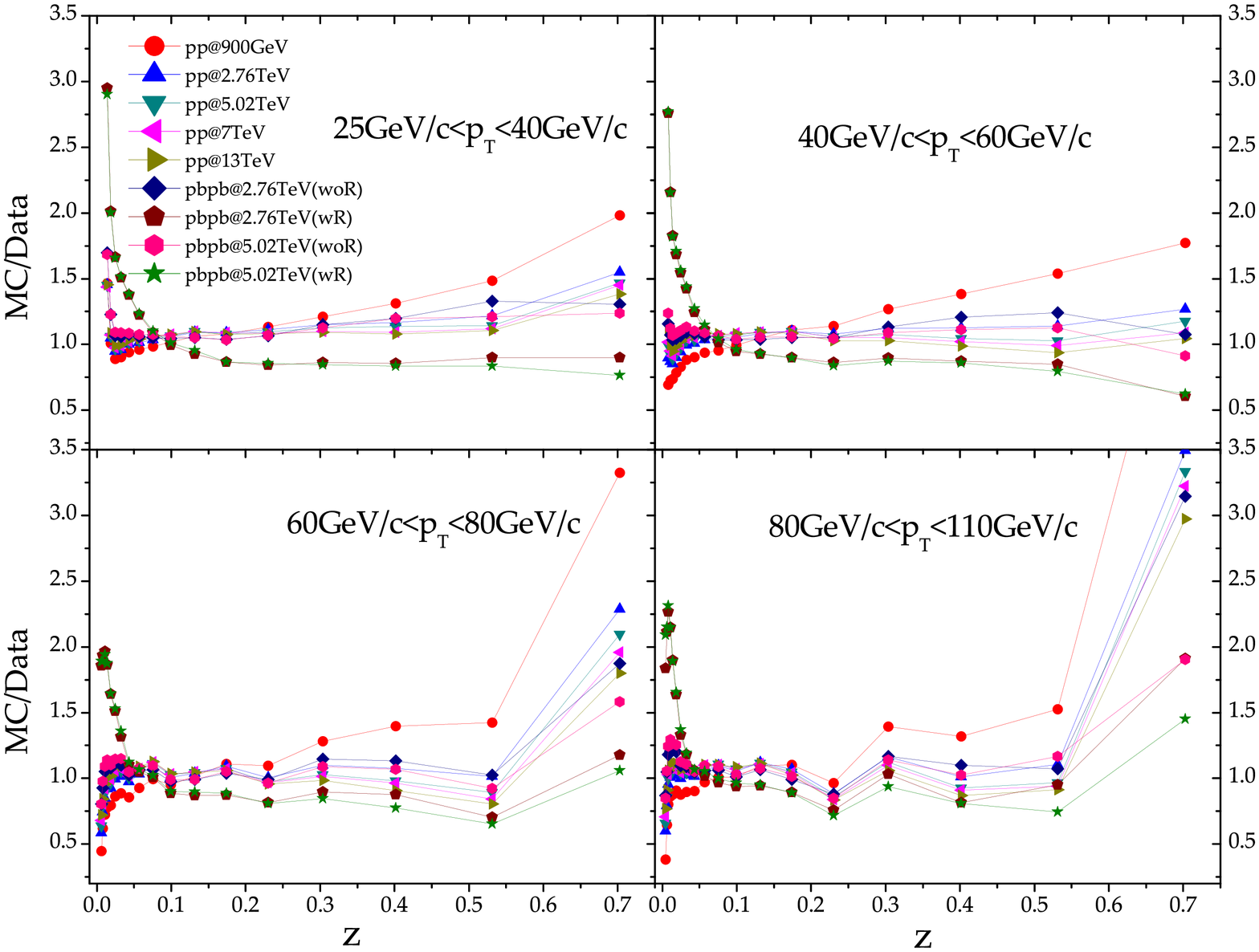}
\figcaption{\label{fig2}   Ratio of simulations of $F(z)$ for $pp$ collisions at $\sqrt{s}$=900 GeV, 2.76 TeV, 5.02 TeV, 7 TeV, 13 TeV and $Pb+Pb$ collisions at $\sqrt{s_{NN}}$=2.76 TeV and 5.02 TeV by JEWEL to data of $\sqrt{s}$=7 TeV $pp$ collisions at four jet momentum intervals.}
\end{center}

\noindent It's obvious that $F(z)$ in Figure \ref{fig2} is suppressed from $Pb+Pb$ collisions to $pp$ collisions, so called $nuclear ~modification ~factor$, which is suppressed about 15\% at $p_{T,jet}\in$[25,40] GeV/c and about 10\% at $p_{T,jet}\in$[80,110] GeV/c at $z>0.1$. At $z<0.1$, the nuclear modification factor is obviously enhanced by a factor 2 to 3 with the decrease of $p_{T,jet}$. For $p_{T,jet}>$60 GeV/c, the suppression goes up at very low $z$ and then goes down at $z<0.01$. For $pp$ collisions, the ratios are all enhanced at low $z<0.02$ and then keeps almost at 1 at large $z$ for $p_{T,jet}<$40 GeV/c, while for high momentum jet at $z<0.02$ the ratio is suppressed up to 50\%. The fragmentation becomes harder for high $p_{T,jet}$ and weak dependence on beam energies and medium effects.

\begin{center}
\includegraphics[width=0.5\textwidth]{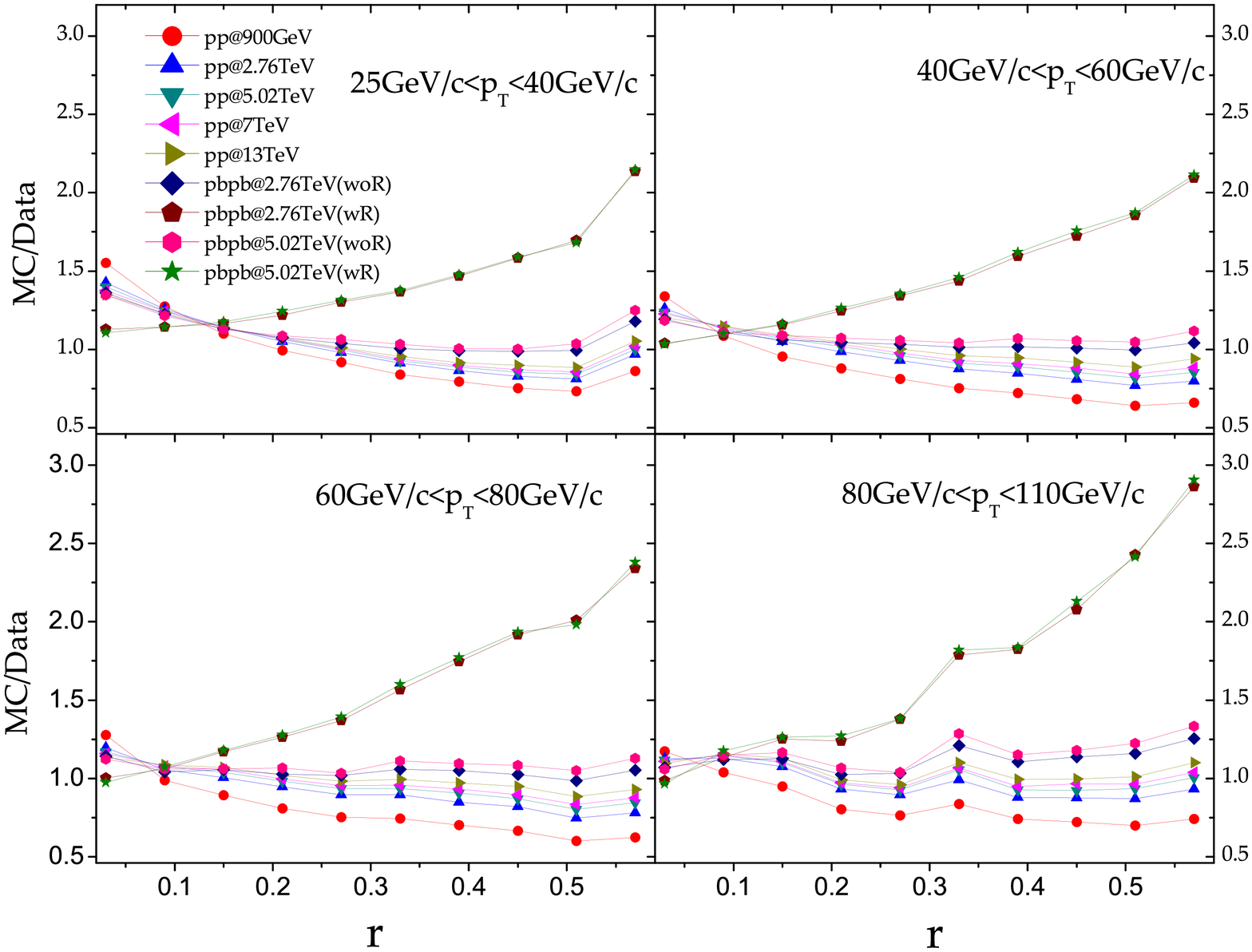}
\figcaption{\label{fig4}   Ratio of simulations of $\rho(r)$ for $pp$ collisions at $\sqrt{s}$=900 GeV, 2.76 TeV, 5.02 TeV, 7 TeV, 13 TeV and $Pb+Pb$ collisions at $\sqrt{s_{NN}}$=2.76 TeV and 5.02 TeV by JEWEL to data of $\sqrt{s}$=7 TeV $pp$ collisions at four jet momentum intervals.}
\end{center}
\vspace{-0.3cm}

\noindent The nuclear modification factor of charged particle density $\rho(r)$ indicates a small suppression in small jet radius and linear increase with the increase of jet radius from $pp$ to $Pb+Pb$ collisions. The slope is about 2, which is the same level of enhancement in $F(z)$. While for the $pp$ collisions, the ratio is almost a constant and slight different with center of mass energy.

\begin{center}
\includegraphics[width=0.5\textwidth]{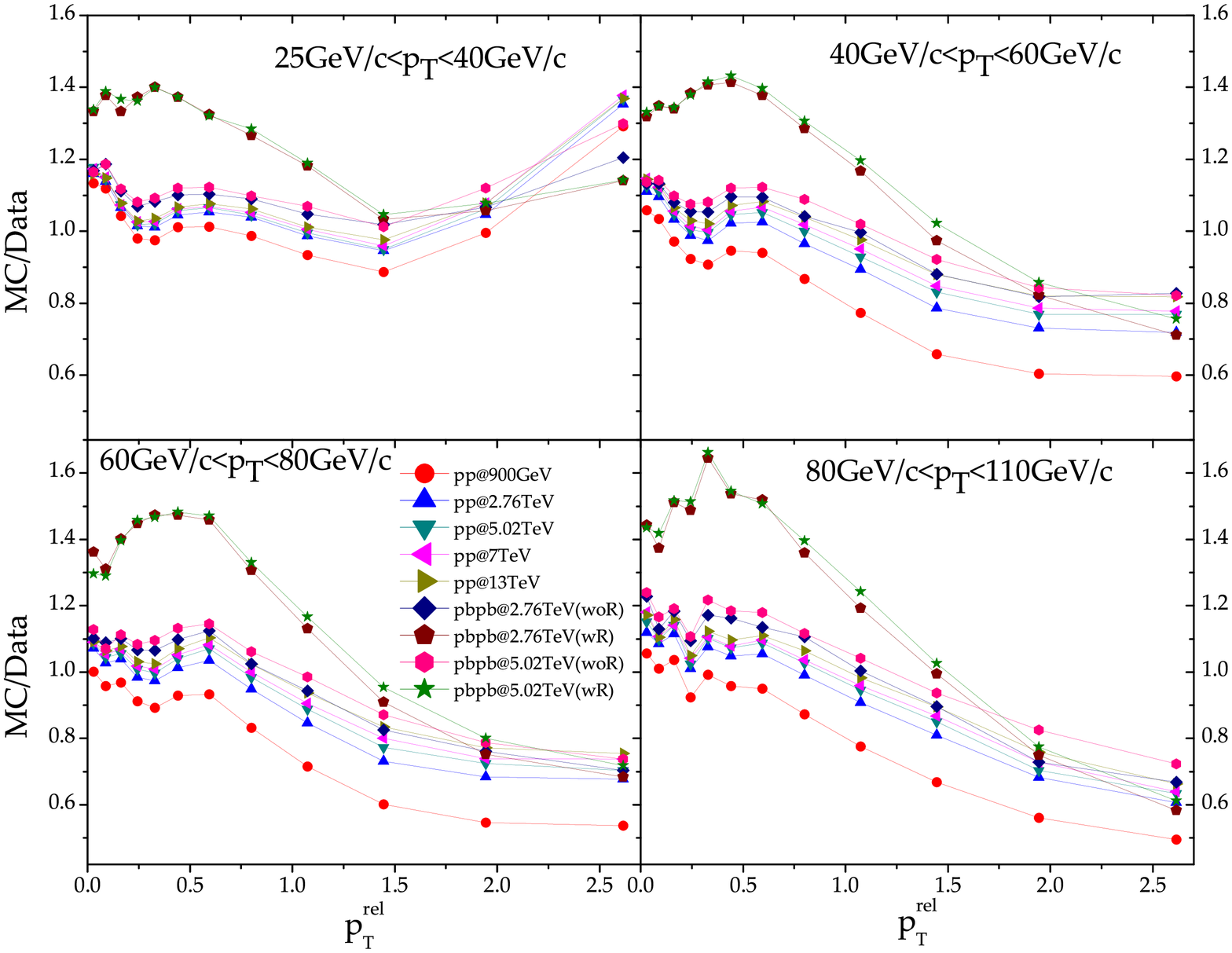}
\figcaption{\label{fig3}  Ratio of simulations of $f(p_{T}^{rel})$ for $pp$ collisions at $\sqrt{s}$=900 GeV, 2.76 TeV, 5.02 TeV, 7 TeV, 13 TeV and $Pb+Pb$ collisions at $\sqrt{s_{NN}}$=2.76 TeV and 5.02 TeV by JEWEL to data of $\sqrt{s}$=7 TeV $pp$ collisions at four jet momentum intervals.}
\end{center}
\vspace{-0.3cm}

\noindent The weak dependence on jet momentum $p_{T,jet}$ and center of mass energies of $pp$ collisions, and strong dependence on medium effect are observed in the ratio plot of $f(p_{T}^{rel})$ in Figure \ref{fig3}. The nuclear modification factor of  $p_{T}^{rel}$ shows an enhancement up to 50\% at low $p_{T}^{rel}<0.5$ and then goes down sharply at larger $p_{T}^{rel}$. At very low $p_{T}^{rel}$ in $pp$ collisions, it seems an fluctuation curve which may indicate flow effect in $pp$ collisions, since $p_{T}^{rel}$ and $\rho_{ch}^{r}$ are directly related to non-perturbative hadronization process controlled by perturbative QCD radiation. With the increase of $p_{T,jet}$, soft gluon radiation is important, which will contribute to jet broadening and mean value of $p_{T}^{rel}$ to rise slowly.

As demonstrated in \cite{smallrALICE}, in this paper, these quantities are also dedicated on study at smaller jet-radius($r=0.1, 0.12, ... ,0.28, 0.30$) to explore the medium effect and jet evolution. The ${girth}$ at small jet-radius $r<0.2$ with different configurations are presented in Figure \ref{fig5}. The trend of simulation is similar with data, although it could not reproduce the data well. The angularity is the first radial moment of jets, which is very sensitive to its radial energy profile. The girth distributions indicate that with the increase of jet-radius, the peak value of $girth$ shifts to large.


 It's interesting to observe that $LeSub$ distribution is exponential decrease with $LeSub$ under different jet-radius cuts. For small jet-radius, asymmetric parton splitting is more probability and symmetric branching could be found at large jet-radius, which is directly related to gluon and quark jet fraction. The momentum dispersion distribution show that $f(p_{T}^{Disp})$ is locally maximum at $P_{T}^{Disp}=0.45$ and then decrease with the increase of jet radius at large $P_{T}^{Disp}$, which means that large energy fraction within smaller jet-radius.

\end{multicols}

\begin{center}
\includegraphics[width=1\textwidth]{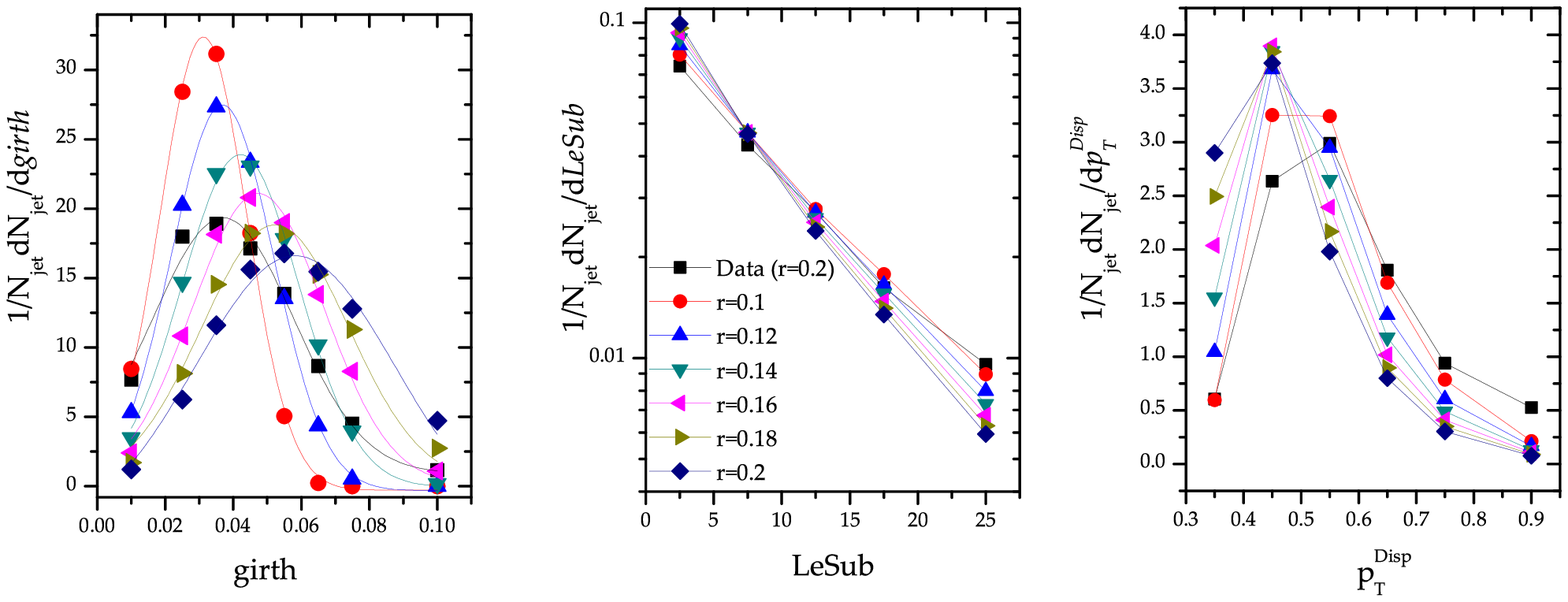}

\figcaption{\label{fig5}   Jet shape distributions of $girth$~(left), $LeSub$~(middle) and $P_{T}^{Disp}$~(right) in 0$\sim$10\% cental $Pb+Pb$ collisions at $\sqrt{S_{NN}}$=2.76 TeV in $p_{T,jet}^{ch}\in$[40,60] GeV/c under different jet-radius cut selections by JEWEL, and the simulations are compared with data\cite{smallrALICE}.}
\end{center}
\vspace{-0.3cm}

\begin{multicols}{2}
\begin{center}
\includegraphics[width=0.45\textwidth]{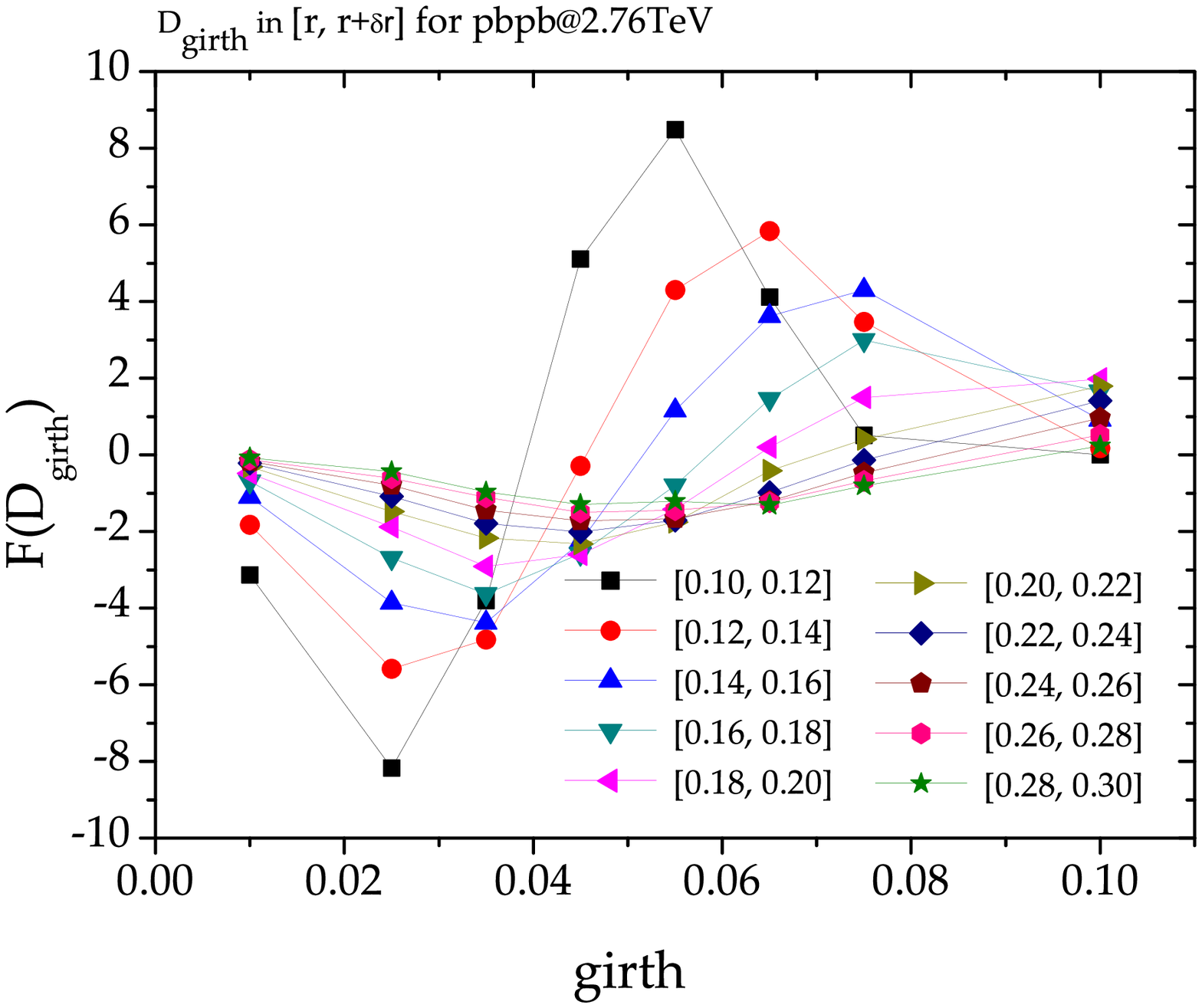}
\figcaption{\label{fig6}   Differential $girth$ distribution in 0$\sim$10\% cental $Pb+Pb$ collisions at $\sqrt{S_{NN}}=$2.76 TeV in $p_{T,jet}^{ch}\in$[40,60] GeV/c at jet radius of $[r,r+\delta r]$ by JEWEL.}
\end{center}
\vspace{-0.3cm}
Figure \ref{fig6} shows the differential $girth$ distribution in 0$\sim$10\% cental $Pb+Pb$ collisions at $\sqrt{S_{NN}}=$2.76 TeV in $p_{T,jet}^{ch}\in$[40,60] GeV/c. It is found that the differential girth evolution from small jet-radius interval to larger behaves as a waveform, which attenuates and becomes flatten at large jet-radius interval. Since small-$r$ jet shows larger jet energy loss and smaller granularity. $D_{girth}$ is sensitive to soft particles or correlated background arising from hot QCD medium. This wave-like pattern and its evolution may potentially to investigate the medium properties, such as coherence effect, medium density and flow effect etc.

The comparison of differential $girth$ distribution between $0\sim10\%$ central $Pb+Pb$ collisions at $\sqrt{S_{NN}}=$2.76 TeV and $pp$ collisions at $\sqrt{s}=$7 TeV in $p_{T,jet}^{ch}\in$[40,60] GeV/c is plotted in Figure \ref{fig7} under three jet-radius intervals. It's obvious that the waveform in $pp$ collision is ahead of $Pb+Pb$ collisions. It could be explained that jet components propagate faster in vacuum than in medium. And this observation has potential in extracting medium parameters and should be further investigation.

\section{Conclusions}

The paper presents a systematical study on jet shape modifications at the LHC energies by a rich set of quantities, $F(z)$, $p_{T}^{rel}$, $\rho_{ch}(r)$, $girth$, $LeSub$ and $p_{T}^{Disp}$ by JEWEL. It concludes that the medium effect is dominant for jet shape modifications, which has weak dependence on center of mass energy. The jet fragmentation is enhanced significantly at very low $z<0.02$ and the fragmented jet constituents are spread to larger jet-radius linearly for $p_{T}^{rel}<1$. The differential girth is proposed and studied. The waveform attenuate phenomena is observed in $f(p_{T}^{rel})$, $girth$ and $D_{girth}$ distributions. The results on the comparison of $D_{girth}$ from $pp$ to $PbPb$ is interesting that the wave pattern in $pp$ collision is ahead of $Pb+Pb$ collisions in small jet-radius intervals, which should an excellent jet shape quantity to explore its relationship to medium density, jet energy loss, coherence effect, flow effect etc.

\begin{center}
\includegraphics[width=0.45\textwidth]{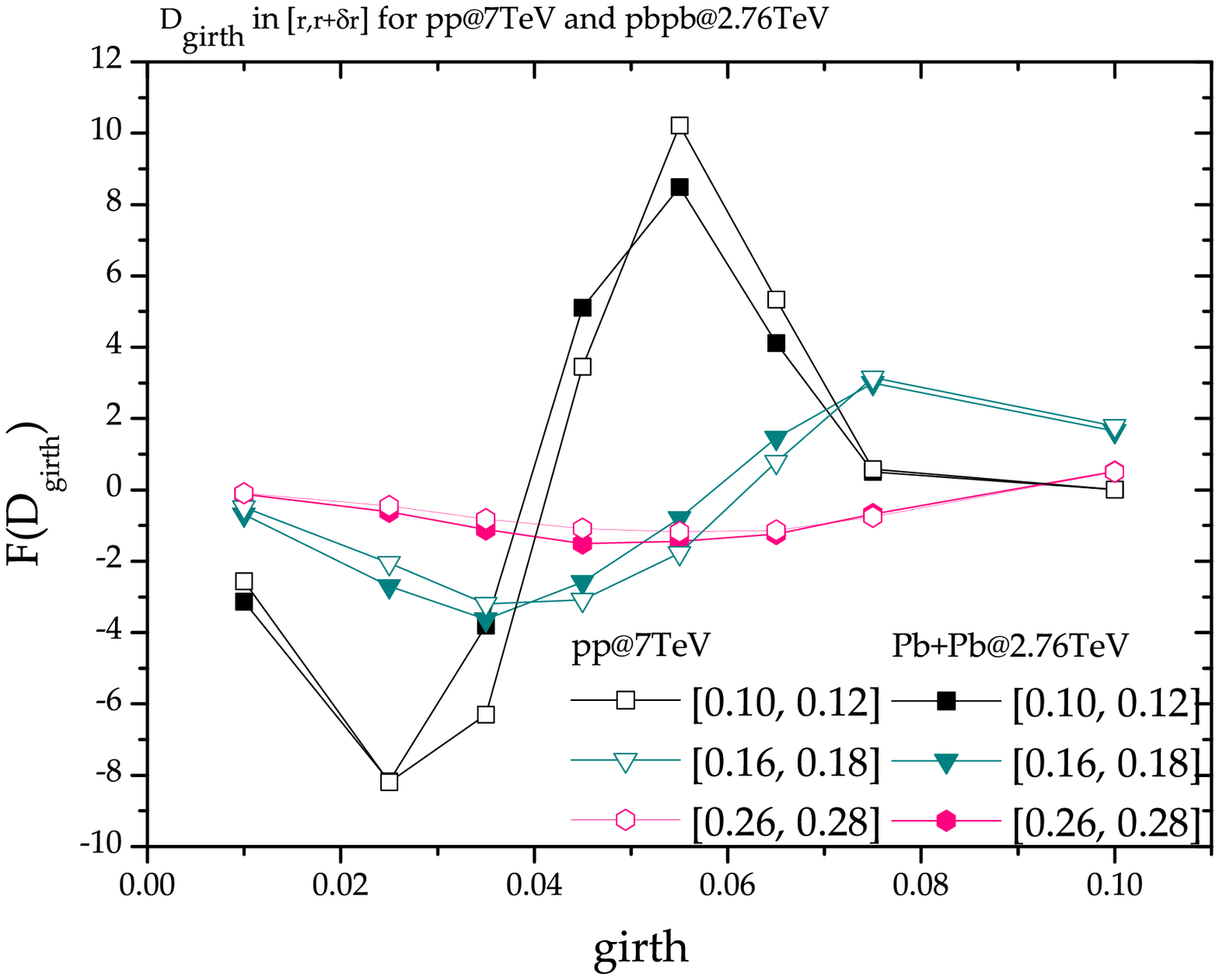}
\figcaption{\label{fig7}   Comparison of differential $girth$ distribution between central $Pb+Pb$ collisions at $\sqrt{S_{NN}}=$2.76 TeV and $pp$ collisions at $\sqrt{s}=$7 TeV in $p_{T,jet}^{ch}\in$[40,60] GeV/c at jet radius of $[r,r+\delta r]$ by JEWEL.}
\end{center}

\section{Acknowledgement}
\acknowledgments{

This work is supported by National Natural Science Foundation of China (grant No. 11505130, 11847014, 11775097, IRG1152106 and 11475068).
}

\vspace{-1mm}
\centerline{\rule{90mm}{0.5pt}}
\vspace{2mm}

\end{multicols}
\end{document}